\documentclass[onecolumn,showpacs,epsfig,prd,preprintnumbers,amsmath,amssymb]{revtex4}
\usepackage{graphicx}
\usepackage{epsfig}
\usepackage{epsf}
\usepackage{dcolumn}
\usepackage{bm}
\newcommand{\be}{\begin{enumerate}}
\newcommand{\im}{\item}
\newcommand{\ee}{\end{enumerate}}

\newcommand{\ccj}{\chi_{c_J}}
\newcommand{\chicj}{\ccj}

\newcommand{\xx}{\Xi^-\bar{\Xi}^+}

\def \jp {J/\psi}

\newcommand{\kz}{K_s^0}
\newcommand{\ppbar}{p\bar p}
\newcommand{\llpp}{\Lambda\bar\Lambda\pi^+\pi^-}
\newcommand{\kkpp}{K_S^0 K_S^0p\bar p}

\newcommand{\psip}{\psi(2S)}
\newcommand{\ks}{K_S^0}

\newcommand{\abcd}[4]{{\vspace*{#1} \hspace*{#2}(a)\hfill (b)\hspace*{#2}}
\vspace*{#3} \\ \hspace*{#2}(c) \hfill(d)\hspace*{#2}\vspace*{#4}}

\begin{document}
\title{\boldmath Measurement of  $\chi_{c_J}$ Decays to $2(\pi^+\pi^-)\ppbar$ Final States}
\author{M.~Ablikim$^{1}$,              J.~Z.~Bai$^{1}$, Y.~Ban$^{12}$,
J.~G.~Bian$^{1}$,              X.~Cai$^{1}$, H.~F.~Chen$^{17}$,
H.~S.~Chen$^{1}$,              H.~X.~Chen$^{1}$, J.~C.~Chen$^{1}$,
Jin~Chen$^{1}$
% ...
\\
\vspace{0.2cm} (BES Collaboration)\\
\vspace{0.2cm} {\it
$^{1}$ Institute of High Energy Physics, Beijing 100049, People's Republic of China\\
% ...
$^{12}$ Peking University, Beijing 100871, People's Republic of China\\
% ...
$^{17}$ University of Science and Technology of China, Hefei 230026, People's Republic of China\\
% ...
} }

\begin{abstract}
Using $\ccj\to2(\pi^+\pi^-)\ppbar$ decays from $14\times 10^6~
\psi(2S)$ events accumulated by the BESII detector at the BEPC, the
intermediate states $\xx$, $\llpp$, and $\kkpp$ are studied, and their
branching ratios or upper limits are measured.
\end{abstract}
\pacs{12.38.Qk, 13.20.Gd, 13.25.Gv, 13.40.Hq}

\maketitle \normalsize

\vspace{1.5cm}

\section{INTRODUCTION}

Experimental studies on charmonia decay properties are essential to
test perturbative QCD models and QCD based calculations. The
importance of the Color Octet Mechanism (COM) for $\ccj$ decays has
been pointed out for many years \cite{octet}, and theoretical
predictions of two-body exclusive decays have been made based on
it. Recently, new experimental results on $\ccj$ exclusive decays have
been reported \cite{bes_chicj,bes}.  COM predictions for many $\ccj$
decays into meson pairs are in agreement with experimental values,
while predictions for some decays into baryon pairs, for example, the
branching fractions of $\ccj\to\Lambda\bar\Lambda$, disagree with
measured values. For further testing of the COM in the
decays of the P-wave charmonia, measurements of other baryon pair
decays of $\ccj$, such as $\ccj\to\xx$ and $\Sigma^0\bar{\Sigma}^0$
are desired.

The measurement of $\chi_{c0}\to\xx$ is helpful for understanding the
Helicity Selection Rule (HSR) \cite{brodsky}, which prohibits $\chi_{c0}$
decays into baryon antibaryon $(B\bar B)$ pairs.
However, the measured branching ratios for $\chi_{c0}$ decays into
$\ppbar$ and $\Lambda\bar\Lambda$ do not vanish, demonstrating a
strong violation of HSR in charmonium decays. Measurements of
$\chi_{c0}$ decays into other baryon anti-baryon pairs would provide
additional tests of the HSR.

In this paper, the analysis of $\psi(2S) \to \gamma
2(\pi^+\pi^-)\ppbar$ decays using $14\times 10^6 ~\psi(2S)$ events
collected at BESII/BEPC is reported. We observe $\ccj \to ~\xx$ and
$\llpp$ decays and search for $\ccj\to\kkpp$; results are compared
with theoretical predictions.

\section{BES DETECTOR}
This analysis is based on a sample of $14\times 10^6$ $\psi(2S)$
events taken with the BESII detector at the BEPC storage ring at a
center-of-mass energy of $M_{\psi(2S)}$ with an integrated luminosity
of $19.72\pm 0.86 \textrm{~pb}^{-1}$ \cite{bes2} . BES is a
conventional solenoidal magnet detector that is described in detail in
Ref. \cite{detector}; BESII is the upgraded version
\cite{detectorii}. A 12-layer vertex chamber (VC) surrounding the beam
pipe provides trigger and trajectory information. A forty-layer main
drift chamber (MDC), located radially outside the VC, provides
trajectory and energy loss (dE/dx) information for charged tracks over
85\% of the total solid angle. The momentum resolution is $\sigma_p/p
= 0.017\sqrt{1 + p^2}$ (p in GeV/c), and the dE/dx resolution for
hadron tracks is $\sim 8\%$. An array of 48 scintillation counters
surrounding the MDC measures the time-of-flight (TOF) of charged
tracks with a resolution of $\sim 200$ ps for hadrons. Radially
outside the TOF system is a 12 radiation length, lead-gas barrel
shower counter (BSC). This measures the energies of electrons and
photons over $\sim 80\%$ of the solid angle with an energy resolution
of $\sigma_E/E = 22\%/\sqrt{E}$ (E in GeV). Outside of the solenoidal
coil, which provides a 0.4 Tesla magnetic field over the tracking
volume, is the iron flux return which is instrumented with three double layers
of counters that identify muons with momentum greater than 0.5 GeV/c.

\section{MONTE-CARLO SIMULATION}
 A GEANT3 based Monte Carlo simulation program, SIMBES \cite{simbes},
 which simulates the detector response, including interactions of
 secondary particles in the detector materials, is used to determine
 detection efficiencies and mass resolutions, as well as to
 optimize selection criteria and estimate backgrounds. Under the
 assumption of a pure E1 transition, the distribution of polar angle
 $\theta$ of the outgoing photon in $\psip\to\gamma\ccj$ decays is
 given by $1+\alpha \textrm{cos}^2\theta$ with $\alpha=1$, $-1/3$, and
 1/13 for $J=0$, 1, and $2$, respectively \cite{e1}. Angular
 distributions of daughter particles for the sequential decays
 $\ccj\to\xx$, $\Xi\to\Lambda\pi$ are simulated based on the transitional
 amplitude information method \cite{amp_mc}. Angular distributions of
 daughter particles from the other multi-body decays are generated
 isotropically in the center-of-mass of $\psip$ or $\ccj$.

\section{EVENT SELECTION}
The selection criteria described below are similar to those used in
previous BES analyses \cite{bes_chicj}.
\subsection{Photon identification}
A shower cluster in the BSC is considered to be a photon candidate, if it
has an energy deposit of more than 50
 MeV, the angle between the nearest charged track and the cluster
is greater than 12$^{\circ}$, the first hit is in the beginning six
radiation lengths, and the difference between the angle of the
cluster development direction in the BSC and the photon emission
direction is less than 37$^{\circ}$.
\subsection{Charged particle identification}
Each charged track is required to have a good helix fit and a
polar angle $\theta$ that satisfies $|\textrm{cos}\theta|<0.8$. The TOF
and $dE/dx$ measurements of the charged track are used to calculate
$\chi^2_{PID}$ values and the corresponding confidence levels for
the hypotheses that the particle is a pion, kaon, or proton.

\subsection{Event selection criteria}

Candidate events are required to satisfy the following selection
criteria:
\be
\im The number of charged tracks is required to be six
with net charge zero, and the number of photon candidates must be less
than four.
\im The  proton and
antiproton must be identified using particle identification;
their particle identification
confidence levels must be greater than
1\%.

\im Four-constraint (4-C)
kinematic fits to the $\psi(2S)\to \gamma 2(\pi^+\pi^-)\ppbar$
hypothesis are performed using each photon candidate. The
combination with the minimum combined
$\chi^2_{comb}=\sum_i^{nchrg}(\chi^2_{dE/dx}+\chi^2_{TOF})_i+\chi^2_{4C}$
is selected, and the  4-C fit confidence level for the selected
combination must be
greater than 1\%.
\im To remove the background channels: $\psip\to
\gamma 2(\pi^+\pi^-)K^+K^-,\gamma3(\pi^+\pi^-)$, and
$2(\pi^+\pi^-)\ppbar$, $\chi^2_{comb}$ of the signal channel is
required to be less than those of the background channels, i.e.,
$\chi^2_{comb}(signal)<\chi^2_{comb}(bg)$.
\ee

\section{DATA ANALYSIS}
\subsection{$\ccj\to\xx$}

\begin{figure}[htbp]
\vspace*{-0.0cm}
\begin{center}
\epsfysize=15cm \epsffile{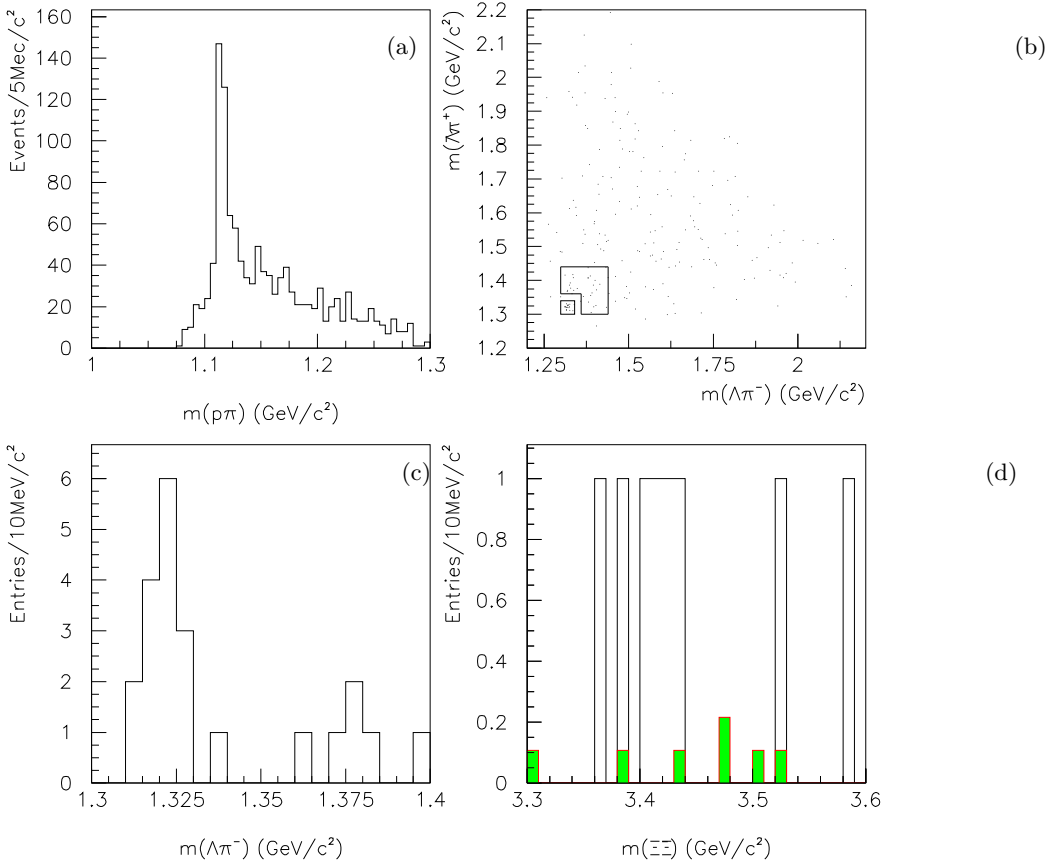}

{\vspace{-13cm} \hspace{4cm}(a)\hfill (b)\hspace{4cm} }
\vspace{5.3cm}\\{\hspace{4.2cm}(c)\hfill  (d)\hspace{4.5cm}
}\vspace{6cm}\\
\parbox{0.8\textwidth}{\caption{(a) The  $p\pi$ invariant
mass distribution. (b) Scatter plot of $m_{\Lambda\pi^-}$
versus $m_{\bar\Lambda\pi^+}$, where the small square corresponds to
the $\xx$ signal region and the larger closed area to the $\xx$
sideband region. (c)  $\Lambda\pi^-$ mass distribution for $|m_{\bar
\Lambda\pi^+}-m_{\Xi}|<17.5$ MeV/c$^2$. (d) The  $m_{\xx}$ invariant mass
distribution, where the shaded histogram corresponds to
the contribution from the $\xx$
sideband region (normalized).\label{data_c2xx}}}
\end{center}
\end{figure}
The sequential decays $\Xi\to\Lambda\pi,\Lambda\to p\pi$ are used to
reconstruct $\ccj\to\xx$. The $p\pi^-$ invariant mass distribution for
candidate events is shown in Figure \ref{data_c2xx}(a), where a clear
$\Lambda$ signal is seen.  A $\bar\Lambda$ signal appears similarly in
the $\bar p\pi^+$ invariant mass distribution. The $\Lambda$ mass
resolution, obtained from MC simulation, is 5.5 MeV/c$^2$.  After
requiring that $\Lambda$ and $\bar\Lambda$ satisfy
$|m_\Lambda-m_{p\pi}|<11$ MeV/c$^2$ $(2\sigma)$, the scatter plot of
$m_{\Lambda\pi^-}$ versus $m_{\bar\Lambda\pi^+}$ is given in Figure
\ref{data_c2xx}(b). In Figure \ref{data_c2xx}(c), the
$m_{\Lambda\pi^-}$ distribution is shown after the requirement
$|m_{\bar\Lambda\pi^+}-m_\Xi|<17.5$ MeV/c$^2$ $(2\sigma)$, where the
$\Xi^-$ appears clearly. After requiring that $\Lambda\pi^-$ and
$\Lambda\pi^+$ satisfy $|M_{\Lambda\pi}-M_\Xi|<17.5$ MeV/c$^2$, the
$m_{\xx}$ distribution is plotted in Figure \ref{data_c2xx}(d), and
the normalized background distribution determined from the
$\Lambda\bar\Lambda$ sideband region is also shown in the plot.

To determine the detection efficiencies, the angular distributions of
the decay particles for the sequential decays
$\psip\to\gamma\ccj,\ccj\to\xx,\Xi\to\Lambda\pi,\Lambda\to p\pi$ are
simulated based on the transitional amplitude information (TAI) method
 \cite{amp_mc}. The detection efficiencies are $(2.3\pm 0.1)\%,
 (2.6\pm 0.1)\%$,  and $(2.3\pm 0.1)\%$
 for $\psip$
sequential decays to $\xx$ via $\chi_{c0}$, $\chi_{c1}$, and $\chi_{c2}$,
respectively.

The main backgrounds from channels with $\Xi$ or $\Lambda$
production, including
$\psip\to\xx,\psip\to\gamma\ccj\to\gamma\Sigma(1385)\bar\Sigma(1385),
\psip\to\gamma\ccj\to\gamma\Sigma^0\bar\Sigma^0$, and
$\psip\to\gamma\ccj\to\gamma\Xi^0\bar\Xi^0$ are determined by
Monte-Carlo simulation. By using the branching ratio of
$\psip\to\xx$ measured in \cite{cleo_bb} and naively assuming
$\ccj\to\Sigma(1385)\bar\Sigma(1385),\Sigma^0\bar\Sigma^0,\Xi^0\bar\Xi^0$
have the same branching ratio as $\ccj\to\Lambda\bar\Lambda$, one
obtains 0.03, 0.01, and 0.02 background events for $\chi_{c0}$,
$\chi_{c1}$, and $\chi_{c2}$, respectively.

\begin{figure}[htbp]
\vspace*{-2.0cm}
\begin{center}
\hspace*{-0.cm} \epsfysize9cm \epsffile{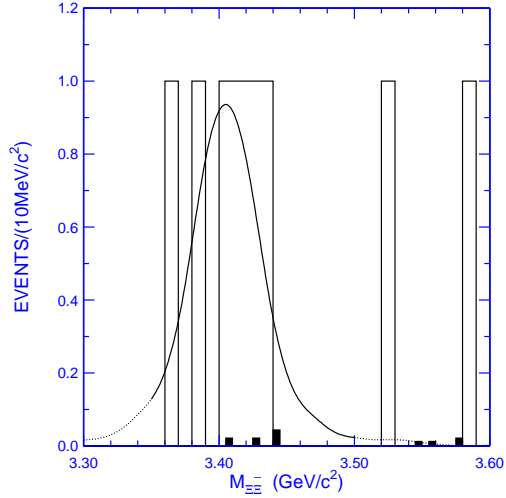} \vspace*{0cm}
\parbox{0.8\textwidth}{\caption{Fit to the $\xx$ mass spectrum for
$\chi_{c0}\to\xx$ with the signal shape obtained from Monte-Carlo
simulation. The histogram corresponds to the data, and
the shaded histogram to the normalized backgrounds obtained from
Monte-Carlo simulation.\label{fitc2xx}}}
\end{center}
\end{figure}

Figure \ref{fitc2xx} shows the fit to the $m_{\xx}$ distribution
using a signal shape obtained from MC simulation. One obtains
$6.4\pm 3.2$ events for $\chi_{c0}\to\xx$. Within twice the mass
resolution of the $\chi_{c1}$ and $\chi_{c2}$, one event is
respectively observed. Upper limits at the $90\%$ Confidence Level
(CL) for $\chi_{c1}\to\xx$ and $\chi_{c2}\to\xx$ are evaluated with
POLE \cite{pole} including systematic uncertainties, and the results
are listed in Table \ref{brc2xx}. The branching ratios and upper
limits ($N^{sig}\to N_{upper}$) are estimated with:
\begin{equation}
B[\chi_{cJ}\to\xx]={N^{sig}/\epsilon \over
N_{\psi(2S)}B[\psi(2S)\to
\gamma\chi_{cJ}]B[\Xi\to\pi\Lambda]^2B[\Lambda\to p\pi]^2},
\end{equation}
where $\epsilon$ denotes the detection efficiency and
$N^{sig}=N^{obs}-N^{bg}$ denotes the number of signal events observed.

\begin{table}[htbp]
\caption{Summary of numbers used in the branching ratio calculation,
where the branching ratio for $\psip\to\gamma\chi_{cJ}$ are taken
from Ref. \cite{cleoc} \label{brc2xx}}
\begin{center}
\begin{tabular}{lccc}
\hline\hline quantity & $\chi_{c0}$  & $\chi_{c1}$  & $\chi_{c2}$
\\\hline
$N^{obs}$ & $6.4\pm 3.2$  & 1.0 & 1.0  \\
$N_{upper}$&12.4&4.6&4.6\\
$\epsilon$ (\%) & $2.3\pm 0.1$  & $2.6\pm 0.1$  & $2.3\pm 0.1$ \\
$N_{\psi(2S)}$ $(\times 10^6)$ & $14.0\pm 0.6$  & $14.0\pm 0.6$  &$14.0\pm 0.6$  \\
$B(\psi(2S)\to\gamma\chi_{cJ})$ (\%) & $9.2\pm 0.5$   & $9.1\pm 0.6$ & $9.3\pm 0.6$  \\
$B(\Xi\to\Lambda\pi)$ (\%) & $99.9\pm 0.1$  & $99.9\pm 0.1$ & $99.9\pm 0.1$ \\
$B(\Lambda\to p\pi)$ (\%) &  $63.9\pm 0.5$ & $63.9\pm 0.5$ &$63.9\pm
0.5$
\\ \hline $B(\chi_{cJ}\to \xx)\times 10^{-4}$ & $5.3\pm 2.7\pm
0.9$ & $<3.4$ (90\% C.L.) &
$<3.7$ (90\% C.L.)  \\
 &  $<10.3$ (90\% C.L.) & & \\
\hline
\end{tabular}
\end{center}
\end{table}

\subsection{$\ccj\to \llpp$}
\begin{figure}[htbp]
\vspace*{-0.0cm}
\begin{center}
\hspace*{-0.cm} \epsfysize15cm \epsffile{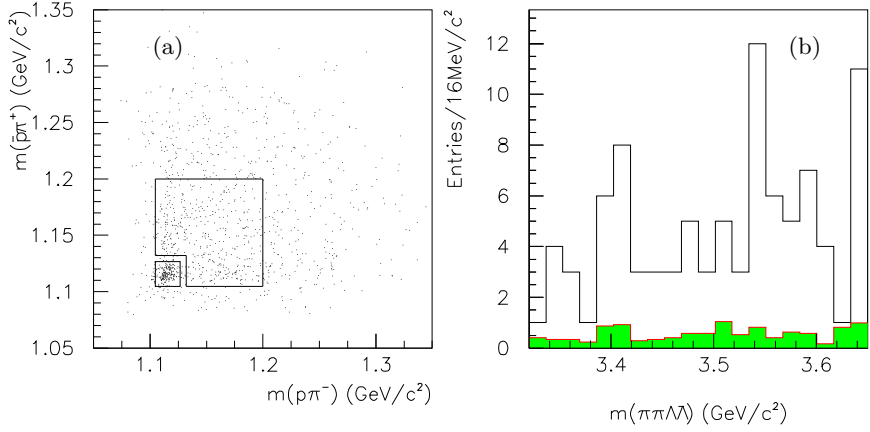}\\
{{\vspace*{-13cm} \hspace*{4cm}(a)\hfill (b)\hspace*{4cm}}
\vspace*{5cm}}
\parbox{0.8\textwidth}{\caption{(a) Scatter plot of $m_{p\pi^-}$ versus
$m_{\bar p\pi^+}$ for candidate events, where the square box corresponds
to the $\Lambda\bar\Lambda$ signal region, and the large box beside it
corresponds to the $\Lambda\bar\Lambda$ sideband region. (b)
The $m_{\Lambda\bar\Lambda\pi\pi}$ invariant mass distribution of events in
the $\Lambda\bar\Lambda$ region (full histogram) and normalized
$\Lambda\bar\Lambda$ sideband background (shaded histogram).
\label{llpp_data}}}
\end{center}
\end{figure}

The $\Lambda$ and $\bar\Lambda$ candidates are reconstructed the same
as in the above analysis. The $\ccj\to\xx$ candidates are
excluded. Figure \ref{llpp_data}(a) shows the scatter plot of
$m_{p\pi^-}$ versus $m_{\bar p\pi^+}$ for candidate events. The dense
cluster in the square, with side twice the $\Lambda$ mass resolution,
corresponds to the $\Lambda\bar\Lambda$ signal. The
$\Lambda\bar\Lambda$ sideband region is shown in the same plot. Figure
\ref{llpp_data}(b) shows the distribution of
$m_{\Lambda\bar\Lambda\pi\pi}$ (full histogram) and
$\Lambda\bar\Lambda$ sideband background (shaded histogram).

To determine the detection efficiency, sequential decays via
$\ccj\to\llpp,\Lambda\to p\pi$ are simulated assuming a pure phase
space distribution. After applying the same selection criteria, one
obtains detection efficiencies of $(0.93\pm 0.04)\%, (1.14\pm
0.04)\%$, and $(1.11\pm 0.04)\%$ for $\chi_{c0}$, $\chi_{c1}$, and
$\chi_{c2}$ decays, respectively.  The decrease in detection
efficiency, compared with the two-body sequential decay via
$\ccj\to\xx$, is due to the difference in the momentum distributions
of the decay particles in $\ccj\to\llpp$.

Potential background contaminations, including
$\psip\to\gamma\ccj\to\gamma\rho^0\Delta^{++}\Delta^{--}\to
\gamma2(\pi^+\pi^-)\ppbar,~\psip\to\pi^+\pi^-\jp$, then $\jp\to
\gamma\eta_c\to\gamma p\pi^+\bar\Delta^{--},\gamma\rho^0\ppbar$, and
$\jp\to\eta\ppbar\to\gamma\pi^+\pi^-\ppbar$, are studied by
Monte-Carlo simulation, and it is found  that these background
contaminations are effectively removed by the $\Lambda\bar\Lambda$
sideband background subtraction.

\begin{figure}[htbp]
\vspace*{-7.5cm}
\begin{center}
\hspace*{-0.cm} \epsfysize15cm \epsffile{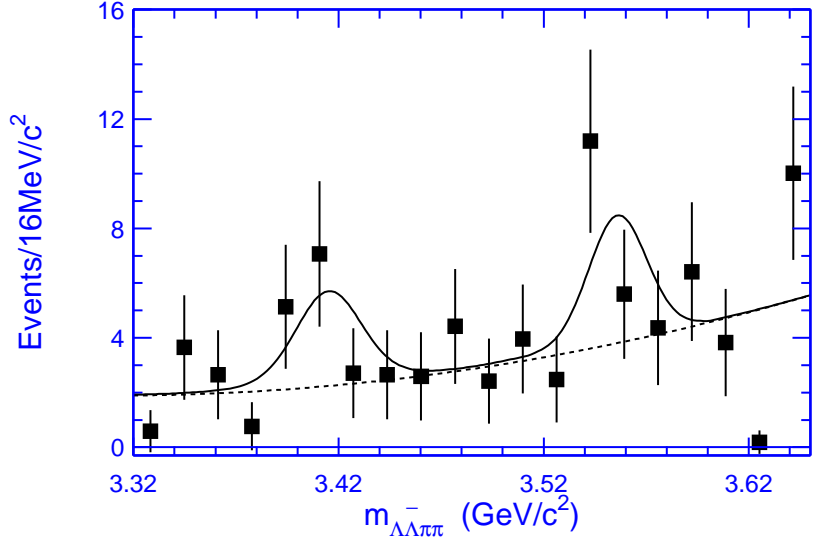}\\
\parbox{0.8\textwidth}{\caption{Fit to the  $m_{\llpp}$ distribution
    after subtracting $\Lambda\bar\Lambda$ sideband background.
The squares with error bars are data, the dashed curve
is a second order polynomial to represent background, and the
solid curve is the fit. \label{llpp_fit}}}
\end{center}
\end{figure}

Figure \ref{llpp_fit} shows the  $m_{\llpp}$ distribution after
subtracting the $\Lambda\bar\Lambda$ sideband background.
Breit-Wigner functions, convoluted with Gaussian mass resolutions,
for the $\chi_{c0},\chi_{c0}$, and $\chi_{c2}$ resonances and a
polynomial background shape are used to fit the $m_{\llpp}$
distribution from 3.3 to 3.65 GeV/c$^2$. The number of signal events
obtained are $n^{sig}=9.6\pm 5.4$, $0.2\pm 3.2$, and $10.4\pm 5.7$
for $\chi_{c0},\chi_{c1}$, and $\chi_{c2}$, respectively. The upper
limits at the 90\% C.L. are estimated with POLE including systematic
uncertainties and also listed in Table~\ref{resllpp}.

\begin{table}[htbp]
\caption{Summary of numbers used in the branching ratio calculation
for $\ccj\to \Lambda\bar\Lambda \pi^+\pi^- $\label{resllpp}.}
\begin{center}
\begin{tabular}{lccc}
\hline\hline quantity & $\chi_{c0}$  & $\chi_{c1}$  & $\chi_{c2}$
\\ \hline
$n^{sig}$ & $9.6\pm 5.4$  & ------ & $10.4\pm 5.7$  \\
$N_{upper}$&$19.6$&$8.8$&$20.4$\\
$\epsilon$ $(\times 10^{-3})$ & $9.3\pm 0.4$  & $11.4\pm 0.4$  & $11.1\pm 0.4$ \\
$N_{\psi(2S)}$ $(\times 10^6)$ & $14\pm 0.6$  & $14\pm 0.6$  & $14\pm 0.6$ \\
$B(\psi(2S)\to\gamma\chi_{cJ})$ (\%) & $9.2\pm 0.5$  & $9.1\pm 0.6$ & $9.3\pm 0.6$  \\
$B(\Lambda\to p\pi)$ (\%) & $63.9\pm 0.5$  & $63.9\pm 0.5$ & $63.9\pm 0.5$ \\
\hline $B(\chi_{cJ}\to \pi^+\pi^-\Lambda\bar\Lambda)\times 10^{-3}$
& $2.0\pm 1.1\pm 0.4$ $(2.5\sigma)$ & ----- &$1.8\pm 1.0\pm 0.3$
 $(2.5\sigma)$  \\
upper-limit $\times 10^{-3}$ (90\% C.L.)&$<4.0$&$<1.5$&$<3.5$ \\
 \hline
\end{tabular}
\end{center}
\end{table}

The branching ratios and upper limits ($N^{sig}\to N_{upper}$) are
estimated using:
\begin{equation}\label{}
B[\chi_{cJ}\to\llpp]={N^{sig}/\epsilon \over
N_{\psi(2S)}B[\psi(2S)\to \gamma\chi_{cJ}]B[\Lambda\to p\pi]^2},
\end{equation}
where $n^{sig}$ denotes the number of signal events. The branching ratios and
upper limits are listed in Table~\ref{resllpp}.

\subsection{$\ccj\to \kkpp$}
$K_S^0$ candidates are selected by choosing the
 $\pi^+\pi^-$ combinations with the minimum value of
$\Delta(i_1,i_2,j_1,j_2)=(m_{\pi^+(i_1)\pi^-(j_1)}-m_{\ks})^2+(m_{\pi^+(i_2)\pi^-(j_2)}-m_{\ks})^2$
 and requiring the $K_S^0$ candidates to satisfy
$|M_{\ks}-M_{\pi^+\pi^-}|<15\textrm{ MeV/c$^2$}~(2\sigma)$. The candidates
for $\ccj\to\llpp$ are excluded.

\begin{figure}[htbp]
\vspace*{0cm}
\begin{center}
\hspace*{-.cm} \epsfysize=13cm \epsffile{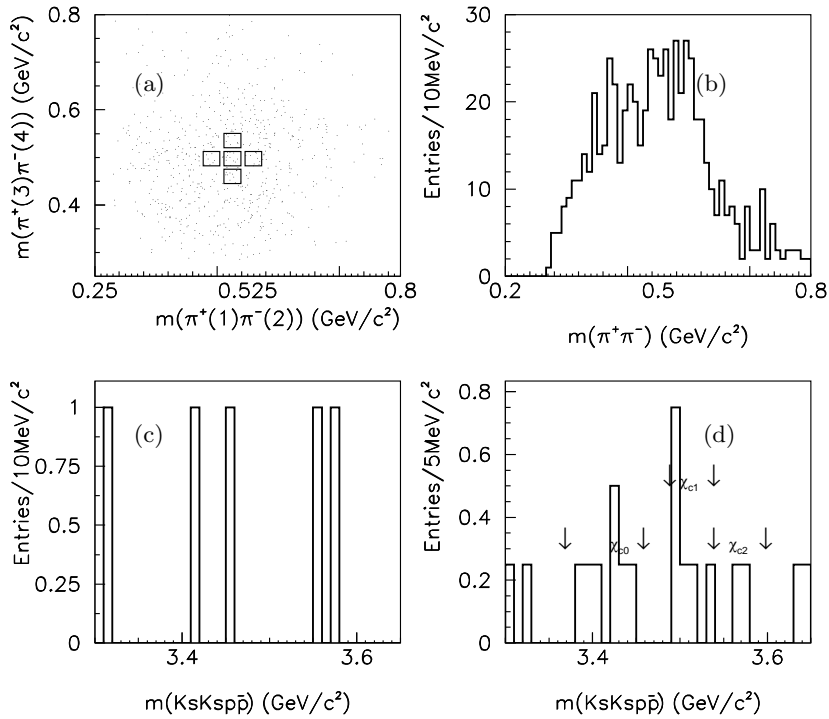}\\
\abcd{-11cm}{4.5cm}{4.3cm}{5cm} \vspace*{-0cm}
\parbox{0.8\textwidth}{ \caption{(a) The scatter plot of $m_{\pi^+\pi^-}$ versus $m_{\pi^+\pi^-}$.
(b) The  $m_{\pi^+\pi^-}$ invariant mass distribution.  (c) The $m_{\ks\ks\ppbar}$
invariant mass distribution. (d) The  $m_{\ks\ks\ppbar}$ invariant
mass distribution (normalized) for events in the $\ks\ks$ sideband regions.
\label{kkpp_data}}}
\end{center}
\end{figure}

The scatter plot of $m_{\pi^+\pi^-}$ versus $m_{\pi^+\pi^-}$ is
shown in Figure \ref{kkpp_data}(a), where the events in the central square
are $\ks\ks$ signal candidates and the four surrounding squares
are the $\ks\ks$ sideband regions.  Figure \ref{kkpp_data}(b) shows
 the $m_{\pi\pi}$ invariant mass distribution. In the $K^0_S$
mass region, no significant signal is observed. After applying the
selection criteria for $K_S^0 K_S^0$,  the distribution of
$m_{\kkpp}$ is given in Figure \ref{kkpp_data}(c). Within selection
regions which are twice the $\ccj$ mass resolution, the observed
numbers of events are 2, 0, and 2 for $\chi_{c0},~\chi_{c1}$, and
$\chi_{c2}$, respectively. Figure \ref{kkpp_data}(d) shows the
background from the $\ks\ks$ sideband background region. The $\ks$
sideband regions are defined as $445.1\textrm{
MeV/c$^2$}<m_{\pi^+\pi^-}<475.1\textrm{ MeV/c$^2$}$ and
$520.1\textrm{ MeV/c$^2$}<m_{\pi^+\pi^-}<550.1\textrm{ MeV/c$^2$}$,
and the mass of other two pions is required to satisfy
$|M_{\ks}-M_{\pi^+\pi^-}|<15\textrm{ MeV/c$^2$}$. The numbers of
background events observed are 1.3, 1.5, and 0.5 for $\chi_{c0}$,
$\chi_{c1}$, and $\chi_{c2}$ respectively. The upper limits at the
90\% C.L. are estimated with POLE including systematic
uncertainties, and the results are listed in Table \ref{upc2kkpp}.

The detection efficiencies are $(0.87\pm 0.06)\%, (0.92\pm 0.06)\%$,
and $(0.90\pm 0.06)\%$ for $\chi_{c0}$, $\chi_{c1}$, and
$\chi_{c2}$ decays, respectively. Upper limits are estimated using
\begin{equation}
B[\chi_{cJ}\to\kkpp]={N_{upper}/\epsilon \over
N_{\psi(2S)}B[\psi(2S)\to \gamma\chi_{cJ}]B[\kz\to \pi^+\pi^-]^2},
\end{equation}
where  $\epsilon$ denotes the detection efficiency. The
upper limits for $\ccj\to\kkpp$ are listed in Table \ref{upc2kkpp}.
\begin{table}[htbp]
\begin{center}
\caption{Summary of numbers used in the branching ratios calculation
of $\ccj\to\kkpp$. \label{upc2kkpp}}
\begin{tabular}{lccc}
\hline\hline quantity & $\chi_{c0}$  & $\chi_{c1}$  & $\chi_{c2}$
\\ \hline
$n_{upper}$ & 4.7  & 2.5 & 4.4  \\
$\epsilon$ $(\times 10^{-3})$ & $8.7\pm 0.6$  & $9.2\pm 0.6$  & $9.0\pm 0.6$ \\
$N_{\psi(2S)}$ $(\times 10^6)$ & $14.0\pm 0.6$   & $14.0\pm 0.6$  & $14.0\pm 0.6$  \\
$B(\psi(2S)\to\gamma\chi_{cJ})$ (\%) & $9.2\pm 0.5$  & $9.1\pm 0.6$ & $9.3\pm 0.6$  \\
$B(\ks\to \pi^+\pi^-)$ (\%) & $69.0\pm 0.1$   & $69.0\pm 0.1$
&$69.0\pm 0.1$
\\ \hline
$B(\chi_{cJ}\to \kkpp)\times 10^{-4}$ & & &  \\
upper-limit (90\% C.L.) &$<8.8$&$<4.5$&$<7.9$ \\
 \hline
\end{tabular}
\end{center}
\end{table}

\section{SYSTEMATIC ERRORS}
Systematic errors on the branching fractions mainly originate from
the MC statistics, the track error matrix, the kinematic fit,
particle identification, the photon efficiency, the uncertainty of
the branching fractions of the intermediate states (from PDG or
published papers), the total number of $\psip$ events, the fitting,
the uncertainty of the angular distribution of the $\Xi$ in $\ccj$
decays, and efficiency corrections from $\ccj\to
\Sigma^-(1385)\bar\Lambda\pi^++c.c.$.

\be
 \im As studied by a previous BES analysis \cite{mdc} using clean
channels like $\jp\to\Lambda\bar\Lambda$ and
$\psi'\to\pi^+\pi^-\jp\to\pi^+\pi^-\mu^+\mu^-$, it is found that the
 tracking efficiency for
MC simulation agrees with data within (1-2)\% for each charged
track. Hence, we take 12\% as the systematic error for events with six
charged tracks.

\im The photon detection efficiency was studied using different
methods with $\jp\to\pi^+\pi^-\pi^0$ events \cite{photon1}, and
the difference between data and MC simulation is about 2\% for
each photon. We take 2\% as the systematic error for decays with
one photon.

\im A kinematic fit is applied to the channels analyzed, and
a probability $>0.01$ is required. The consistency between data and
Monte Carlo efficiencies for the kinematic fit depends on whether the
track fitting error matrices are consistent for data and MC
simulation. From earlier studies, we take 4\% as the systematic error
for kinematic fitting \cite{4cfit}.

\im The $p$ and $\bar{p}$ candidates are identified with the
requirement that the particle identification confidence level be
greater than $0.01$. Comparing data and MC data samples for the
decay $\psip\to\pi^+\pi^-\jp,\jp\to\pi^0\ppbar$, the efficiency
difference is about 2.8\% for one particle identified as a proton or
antiproton. Here 5.6\% is taken as the systematic error for $\ppbar$
identification.

\im To estimate the background uncertainty for $\ccj\to\xx$,
background MC
samples and $\xx$ sideband backgrounds are used. The maximum difference
between them is about 5\%, which is taken as the systematic error. For
$\ccj\to\llpp$, the background uncertainty includes the uncertainty in
the background
shape and fitting interval; the differences found using different
shapes and intervals are 5.25\%,
3.57\%, and 2.85\% for $\chi_{c0}$, $\chi_{c1}$, and $\chi_{c2}$ decays,
respectively. For $\ccj\to\kkpp$ decays, the maximum difference in
estimation of the upper limits between different choices of $\ks\ks$
sideband regions is about 4\%, which is taken as the systematic error.

\im To determine the detection efficiency for $\ccj\to \xx$, the TAI
method is used to account for the angular distribution of the $\Xi$.
The uncertainty of the parameters used introduce 4\% and 5\%
differences in efficiencies of the $\chi_{c1}$ and $\chi_{c2}$
channels, respectively, which are taken as the systematic error for
the uncertainty of the $\Xi$ angular distribution.

\im In the analysis of $\ccj\to\llpp$, the $\xx$ intermediate state
is excluded, and no significant signals for decays
$\ccj\to\Sigma^+(1385)\bar\Sigma^-(1385),\Lambda(1405)\bar\Lambda(1405)$
are observed. The intermediate state corrections to the detection
efficiency are evaluated by considering only the observable decay
$\ccj\to\Sigma^-(1385)\bar\Lambda\pi^++c.c.$. The branching
fractions are roughly estimated using data. The efficiency
differences are evaluated by:
$${\delta\epsilon\over \epsilon_1}={|\epsilon_1-\epsilon_2|\over \epsilon_1}{B(\ccj\to\Sigma^-(1385)\bar\Lambda \pi^++c.c.)B(\Sigma^-\to\Lambda\pi^-)\over
B(\ccj\to\Sigma^-(1385)\bar\Lambda
\pi^++c.c.)B(\Sigma^-\to\Lambda\pi^-)+B(\ccj\to\llpp)},$$  where
$\epsilon_1$ and $\epsilon_2$ are the detection efficiencies for
$\psip$ radiative decays via $\ccj\to\llpp$ and
$\ccj\to\Sigma^-(1385)\bar\Lambda\pi^++c.c.,\Sigma(1385)\to\Lambda\pi$,
respectively. We obtain the efficiency difference $\delta
\epsilon/\epsilon_1=7.1\%, 6.0\%$, and 2.6\% for decays of
$\chi_{c0}$, $\chi_{c1}$, and $\chi_{c2}$, respectively, which is
treated as one of the systematic errors. Total systematic errors are
given in Table \ref{syserr}. \ee

\begin{table}[htbp]
\caption{Summary of systematic errors (\%).\label{syserr}}
\begin{center}
\begin{tabular}{l|ccccccccccc}
\hline\hline\\
 Source& \multicolumn{3}{c}{$\ccj\to\xx$}& &
\multicolumn{3}{c}{$\ccj\to\llpp$}& &
\multicolumn{3}{c}{$\ccj\to\kkpp$}\\\cline{2-4}\cline{6-8}\cline{10-12}
&$\chi_{c0}$&$\chi_{c1}$&$\chi_{c2}$&
&$\chi_{c0}$&$\chi_{c1}$&$\chi_{c2}$&&$\chi_{c0}$&$\chi_{c0}$&$\chi_{c2}$\\\hline
MDC track& 12.0&12.0&12.0&& 12.0&12.0&12.0&& 12.0&12.0&12.0\\
Photon efficiency &2.0&2.0&2.0& &2.0&2.0&2.0& &2.0&2.0&2.0\\
4C-fit&4.0&4.0&4.0& &4.0&4.0&4.0& &4.0&4.0&4.0\\
Number of $\psip$ &4.0&4.0&4.0& &4.0&4.0&4.0& &4.0&4.0&4.0\\
$\ppbar$ particle ID&5.6&5.6&5.6& &5.6&5.6&5.6& &5.6&5.6&5.6\\
$B(\Lambda\to p\pi)$&0.8&0.8&0.8& &0.8&0.8&0.8& &0.8&0.8&0.8\\
$B(\psip\to \gamma\ccj)$&5.1&6.4&6.7& &5.1&6.4&6.7& &5.1&6.4&6.7\\
Background uncertainty&5.0&5.0&5.0& &5.3&3.6&2.9&&4.0&4.0&4.0\\
MC statistics &2.0&1.9&2.1&&3.5&3.0&3.0&&5.0&5.0&5.0\\
Angular distribution of $\xx$&--&4.0&5.0&&--&--&--&&--&--&--\\
Intermediate state
$\Sigma(1385)$&--&--&--&&7.1&6.0&2.6&&--&--&--\\\hline
 Total &16.3&17.2&17.6&&18.1&17.6&16.7&&16.7&17.1&17.3\\\hline\hline

\end{tabular}
\end{center}
\end{table}
\section{RESULTS AND DISCUSSION}
Using the $14\times10^6~ \psip$ events accumulated at BESII, the
analysis of $\ccj\to\xx,\llpp,\kkpp$ is carried out. The measured
branching fractions or upper limits are summarized in Table
\ref{sumtab}, along with some theoretical predictions. Theoretically,
the quark creation model (QCM) predicts $B(\chi_{c0}\to\xx)=(2.3\pm
0.7)\times 10^{-4}$, which is consistent with the experimental value
within $1\sigma$. For $\chi_{c1}$ and $\chi_{c2}$ decays into $\xx$,
the measured upper limits cover both the COM and QCM predictions.

Within $1.8\sigma$ the branching fraction of $\chi_{c0}\to\xx$ does
not vanish. For further testing of the violation of the HSR in this
decay, higher accuracy measurements are required.

\begin{table}[htbp]
\begin{center}
\parbox{0.8\textwidth}{\caption{The comparison of the branching fractions or upper
limits for $\chicj\to \xx$, $\llpp$, and $\kkpp$ between experimental
values and theoretical predictions. The COM predictions are
from Ref. \cite{com_pre}, and the quark creation model (QCM) predictions
are from Ref. \cite{qcm_pre}. \label{sumtab}}}
\begin{tabular}{llcc}
\hline\hline \\Decay modes & Branching ratios&
\multicolumn{2}{c}{Theoretical predictions}\\\cline{3-4} & &COM
&QCM
\\\hline
$\chi_{c0}\to\xx$ & $(5.3\pm 2.7\pm 0.9)\times 10^{-4}$&-----&$(2.3\pm 0.7)\times 10^{-4}$  \\
&or $<10.3\times 10^{-4}$ (90\% C.L.)&&\\
$\chi_{c1}\to\xx$ & $<3.4\times 10^{-4}$ (90\% C.L.) &$2.4\times 10^{-5}$&-----\\
$\chi_{c2}\to\xx$ & $<3.7\times 10^{-4}$ (90\% C.L.)&$3.4\times
10^{-5}$ &$(4.8\pm 2.1)\times 10^{-5}$\\\hline
$\chi_{c0}\to\llpp$ & $(2.0\pm 1.1\pm 0.4)\times 10^{-3}$ $(2.5\sigma)$ &-----&-----\\
&or $<4.0\times 10^{-3}$ (90\% C.L.) &&\\
$\chi_{c1}\to\llpp$ & $<1.5\times 10^{-3}$ (90\% C.L.) &-----&-----\\
$\chi_{c2}\to\llpp$ & $(1.8\pm 1.0\pm 0.3)\times 10^{-3}$ $(2.5\sigma)$
& &\\ &or $<3.5\times 10^{-3}$(90\% C.L.)&&\\\hline
$\chi_{c0}\to\kkpp$ & $<8.8\times 10^{-4}$ (90\% C.L.)&-----&----- \\
$\chi_{c1}\to\kkpp$ & $<4.5\times 10^{-4}$ (90\% C.L.) &-----&-----\\
$\chi_{c2}\to\kkpp$ & $<7.9\times 10^{-4}$ (90\% C.L.)&-----&-----\\
\hline\hline
\end{tabular}
\end{center}
\end{table}

\section{Acknowledgment}

The BES collaboration thanks the staff of BEPC for their hard
efforts. This work is supported in part by the National Natural
Science Foundation of China under contracts Nos. 10491300, 10225524,
10225525, 10425523, the Chinese Academy of Sciences under contract
No. KJ 95T-03, the 100 Talents Program of CAS under Contract Nos.
U-11, U-24, U-25, and the Knowledge Innovation Project of CAS under
Contract Nos. U-602, U-34 (IHEP), the National Natural Science
Foundation of China under Contract No. 10225522 (Tsinghua
University), and the Department of Energy under Contract
No.DE-FG02-04ER41291 (U Hawaii).

\end{document}